\documentclass[preprint,showkeys,unsortedaddress,superscriptaddress, amsfonts,showpacs,amsmath,amssymb]{revtex4}
\usepackage[dvips]{color}
\usepackage{epsfig}
\usepackage{amsmath}
\usepackage{graphicx}
\begin{document}
\title{Logamediate Inflation by Tachyon Field}

\author{A. Ravanpak}
\email{a.ravanpak@vru.ac.ir} \affiliation{Department of Physics,
Vali-e-Asr University, Rafsanjan, Iran}
\author{F. Salmeh}
\email{fahimehsalmeh@gmail.com} \affiliation{Department of Physics,
Vali-e-Asr University, Rafsanjan, Iran}

\date{\small {\today}}

\begin{abstract}
A logamediate inflationary model in the presence of the tachyon scalar field will be studied. Considering slow-roll inflation, the equations of motion of the universe and the tachyon field will be derived. In the context of perturbation theory, some important perturbation parameters will be obtained and using numerical calculations the consistency of our model with observational data will be illustrated.
\end{abstract}

\maketitle


\section{Introduction}

After encountering some serious problems, cosmologists had to improve the standard big-bang model adding some parts to it. The flatness and horizon problems were the most famous of those problems and the best part which added to the standard model and completed it was the inflationary scenario. Inflation is a short period at very early stages of the history of the universe in which the universe experiences a very rapidly accelerated expansion and the scale factor parameter $a(t)$ grows by many order of magnitude. In terms of how the scale factor varies with time one can classify inflationary models to for example power law inflation $a(t)=t^q$, exponential inflation $a(t)=\exp(pt)$ and intermediate inflation $a(t)=\exp(pt^q)$. Among different models of inflation, the one has not been investigated greatly is the logamediate inflation in which for $t > 1$ the scale factor behaves as
\begin{equation}\label{scale}
a(t)=\exp(A(\ln t)^{\lambda})
\end{equation}
where $A > 0$ and $\lambda > 1$ are constants. One can check that for the case $\lambda=1$ the model reduces to the power law inflation. The logamediate inflation can be extracted from some scalar-tensor theories which naturally give rise these solutions \cite{Barrow1} and also from a new class of cosmological solutions with indefinite expansion which result when weak general conditions apply on the cosmological models \cite{Barrow}. Although these models belong to a class of models called non-oscillating models that can not naturally bring inflation to an end, but different approaches such as curvaton scenario can be used to do this duty \cite{Campo2}.

On the other hand, the kind of the scalar field which plays the role of the inflaton field is also important. The standard scalar field is the most usual one, but some other fields can also be responsible for it. Among them, the tachyon field is of particular interest \cite{Sen2},\cite{Sen3}. Its equation of state parameter varies between 0 and -1 and thus it can be a good choice for the inflaton field \cite{Mazumdar}--\cite{Sami2}. Also it has been shown that the tachyon field can play the role of dark sectors of the universe \cite{Padmanabhan}--\cite{Farajollahi} and even at the same time derive inflation and then behave as dark matter or non-relativistic fluid \cite{Gibbons}. Tachyonic inflation is a special case of k-inflationary models in which the inflaton field starts its evolution from an unstable maximum when $\phi\rightarrow0$ and finally approaches zero when $\phi\rightarrow\infty$.

The concept of logamediate inflation with tachyon field or without it has been analyzed in the literature. For instance, in \cite{Barrow2} the dynamics of the logamediate inflation in the presence of a standard scalar field and its consistency with observational results has been shown in details. In \cite{Herrera}, the authors have been investigated the warm-logamediate inflationary universe in both weak and strong dissipative regimes and obtained the general conditions which are necessary that the model to be realizable. Also, in \cite{Setare}, the warm-logamediate inflation in the presence of the tachyon field as the inflaton has been analyzed only in high dissipative regime and the results have been compared by the observations.

In this work we are trying to use the tachyon field as the inflaton in the logamediate inflationary scenario. Our aim is to obtain the influence of the tachyon field on logamediate inflation in comparison with \cite{Barrow2} and also to fill the gap between the articles noted above. In the next section, we will apply tachyon field in a logamediate inflationary model under slow-roll conditions. Section III. deals with perturbation theory. At this section we will calculate all the perturbation parameters which are needed to have a comparison with observations. The numerical comparisons have been done in the subsection A. Section IV deals with how realistic is our model. At the end, there is a conclusion section in which we will discuss our results.

\section{Logamediate Inflationary Model}

We start with the field equations in a flat Friedmann-Robertson-Walker (FRW) universe
\begin{equation}\label{fried}
3H^{2}=\rho_\phi
\end{equation}
and
\begin{equation}\label{acceleration}
2\dot H+3H^{2}=-p_\phi,
\end{equation}
in which $H=\dot a/a$ is the Hubble parameter, $a = a(t)$ is the scale factor and the dot means derivative with respect to the cosmological time $t$. Here, we have used units such that $8\pi G=c=\hbar=1$. Also, we assume the matter content of the universe is a scalar field, $\phi(t)$, so-called inflaton where $\rho_\phi$ and $p_\phi$ represent its energy density and pressure, respectively and they satisfy the following conservation equation
\begin{equation}\label{conservation}
\dot\rho_{\phi}+3H(\rho_{\phi}+p_{\phi})=0.
\end{equation}

From now on we consider the inflaton field as a tachyon field where its energy density and pressure given by
\begin{equation}\label{rophi}
\rho_{\phi}=\frac{V({\phi})}{\sqrt{1-\dot\phi^{2}}},\quad p_{\phi}=-V({\phi})\sqrt{1-\dot\phi^{2}},
\end{equation}
where $V({\phi})$ is the tachyonic scalar potential. Substituting (\ref{rophi}) in (\ref{conservation}) we reach to the equation of motion of the tachyon field
\begin{equation}\label{field}
\frac{\ddot\phi}{1-\dot\phi^{2}}+3H\dot\phi+\frac{V'}{V}=0,
\end{equation}
where $V^\prime=\partial V(\phi)/\partial\phi$. Also, using (\ref{fried}), (\ref{conservation}) and (\ref{rophi}), one can obtain
\begin{equation}\label{phidot}
\dot\phi=\sqrt{-\frac{2\dot H}{3H^2}}.
\end{equation}

Considering the logamediate inflationary model in which the scale factor $a(t)$ behaves as (\ref{scale}), one can obtain the exact solution of (\ref{phidot}) as
\begin{equation}\label{phi}
\phi(t)=\int\sqrt{\frac{2}{3A\lambda}}(\ln t)^{-\lambda/2}(\ln t -\lambda+1)^{1/2}dt
\end{equation}
Also using (\ref{fried}), (\ref{rophi}) and (\ref{phidot}) the potential of the inflaton field can be obtained as
\begin{equation}\label{v(t)}
V(t)=3(\frac {A\lambda}{t})^2(\ln t)^{2(\lambda-1)}\left(1+\frac{2(\ln t)^{-\lambda}(\lambda-1-\ln t)}{3A\lambda}\right)^{\frac{1}{2}}
\end{equation}

To have a long enough period of inflation we need to our inflaton field rolls slowly down its potential. In this scenario which is called slow-roll inflation the energy density of the inflaton field and its potential satisfy $\rho_{\phi}\sim V$. Thus in our model, under slow-roll conditions i.e. $\dot\phi^2\ll1$ and $\ddot\phi\ll3H\dot\phi$, equations (\ref{fried}) and (\ref{field}) reduce to
\begin{equation}\label{efffried}
3H^2\approx V
\end{equation}
and
\begin{equation}\label{efffield}
\frac{V'}{V}\approx -3H\dot\phi,
\end{equation}
respectively. Also, the tachyonic potential (\ref{v(t)}) becomes
\begin{equation}\label{effv(t)}
V(t)=3(\frac {A\lambda}{t})^2(\ln t)^{2(\lambda-1)}
\end{equation}
There are a few dimensionless parameters in slow-roll inflationary models called slow-roll parameters. In terms of our model parameters they can be written as
\begin{equation}\label{epsilon}
\varepsilon=-\frac{\dot H}{H^2}=\frac{(\ln t)^{-\lambda}}{A\lambda}(\ln t-\lambda+1)
\end{equation}
and
\begin{equation}\label{eta}
\eta=-\frac{\ddot\phi}{H\dot\phi}=\frac{1}{2H}[-\frac{\ddot V}{\dot V}+\frac{\dot H}{H}+\frac{\dot V}{V}].
\end{equation}
One can check that the slow-roll parameter $\varepsilon$ starts to increase at $t=1$, reaches to a maximum at some value $t_\varepsilon$ and then returns and approaches zero as $t\rightarrow\infty$. If we pay attention to those cases in which the maximum value of $\varepsilon$ is greater than one, we can choose $\varepsilon=1$ as the beginning condition of inflation. For these cases ($\varepsilon_{max}>1$), one can obtain a constraint for our model parameters as below
\begin{equation}\label{Alambda}
A<\lambda^{-\lambda-1}
\end{equation}

We can also obtain the number of e-folds between two different values
$t_1$ and $t_2>t_1$ for this model as
\begin{equation}\label{nume-fold}
N=\int_{t_1}^{t_2}H dt=A[(\ln t_2)^\lambda-(\ln t_1)^\lambda]
\end{equation}
where $t_1$ represents the time in which inflation begins.

\section{PERTURBATION}

Although studying a homogeneous and isotropic universe model is sometimes very useful, we know that in a real cosmology there are deviations from homogeneity and isotropic assumptions. This motivates us to investigate the perturbation theory in our model. We believe that inhomogeneities grow with time due to the attractive feature of gravity and thus we can say that they were very smaller in the past. Because of the smallness of them we can use linear perturbation theory. But as it appears from Einstein's equations and to have a more realistic investigation we need a relativistic perturbation theory, i.e. a perturbed inflaton field in a perturbed geometry. So we start by the most general linearly perturbed flat FRW metric which includes both scalar and tensor perturbations as below
\begin{equation}\label{p-metric}
    ds^2 = -(1+2C)dt^2+2a(t)D_{,i}dx^idt+a(t)^2[(1-2\psi)\delta_{ij}+2E_{,i,j}+2h_{ij}]dx^idx^j,
\end{equation}
where $C$, $D$, $\psi$ and $E$ are the scalar metric perturbation and $h_{ij}$ is the transverse-traceless tensor perturbation. A very useful quantity in characterizing the properties of the perturbations, is the power spectrum. First of all, we calculate the power spectrum of the curvature perturbation ${\cal P}_{\cal R}$, which appears in deriving the correlation function of the inflaton field in the vacuum state. For the tachyon field this parameter is defined as
\begin{equation}\label{prtachyon}
{\cal P}_{\cal R}=(\frac{H^2}{2\pi\dot\phi})^2\frac{1}{Z_s}
\end{equation}
where $Z_s=V(1-\dot\phi^2)^{-3/2}$ \cite{Hwang}. Applying slow-roll approximation in (\ref{prtachyon}) and using equations (\ref{efffried}) and (\ref{efffield}), one can obtain
\begin{equation}\label{power-scalarper}
{\cal P}_{\cal R}\approx(\frac{H^2}{2\pi\dot\phi})^2\frac{1}{V}=\frac{-3H^5}{4\pi^2\dot V}.
\end{equation}

When someone deals with perturbation in cosmology a few special parameters have to be identified. The first one is the scalar spectral index $n_s$ which is related to the scalar power spectrum via the relation $n_s-1=d\ln {\cal P}_{\cal R}/d\ln k$, where $d\ln k=dN=Hdt$ \cite{Campo}. With attention to definition of ${\cal P}_{\cal R}$ in the slow-roll approximation, we reach to
\begin{equation}\label{ns}
n_s\approx1+\frac{5\dot H}{H^2}-\frac{\ddot V}{H\dot V}=1+2(\eta-\varepsilon).
\end{equation}
The second interesting parameter is the running in the scalar spectral index parameter $n_{run}$, which has been indicated by one-year to seven-years data set of the Wilkinson Microwave Anisotropy Probe (WMAP) and can be obtained via $n_{run}=d n_s/d \ln k$. Thus with attention to (\ref {ns}) one can reach to equation below
\begin{equation}\label{nrun}
n_{run}\approx\frac{2}{H}(\dot\eta-\dot\varepsilon)
\end{equation}

So far we have only studied the scalar perturbations. But how about tensor contributions? In fact the primordial gravitational waves are these tensor perturbations where are essentially equivalent to two massless scalar fields. Thus, the power spectrum of tensor perturbations can be written as
\begin{equation}\label{power-tensorper}
{\cal P}_g=8(\frac{H}{2\pi})^2
\end{equation}

The third special parameter we deal with is the tensor to scalar
ratio $r$ which by definition and using equations (\ref{power-scalarper}) and (\ref{power-tensorper}) becomes
\begin{equation}\label{r}
r=\frac{{\cal P}_g}{{\cal P}_{\cal R}}\approx16\varepsilon
\end{equation}

\subsection{numerical discussion}

\begin{figure}[h]
\centering
\includegraphics[width=0.48\textwidth]{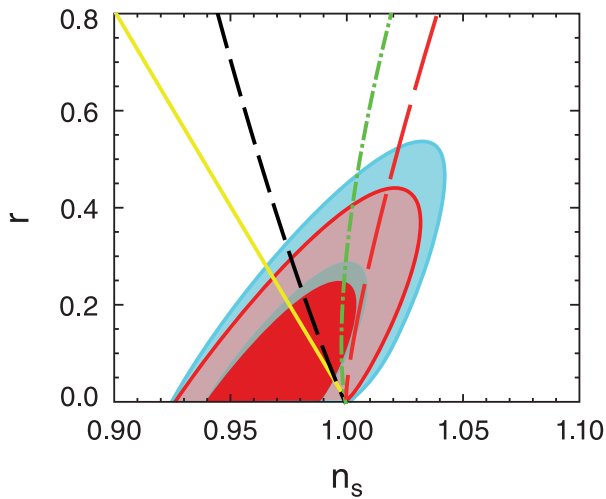}
\caption{The trajectories $r-n_s$ for different combinations of ($\lambda$, A). They have been compared with the five-years (blue regions) and seven years (red regions) data set of WMAP. In each case the contours show 68\% and 95\% confidence regions \cite{Larson}. The solid yellow, dash black, dash-dot green and long dash red curves are related to the combinations (2, 10$^{-1}$), (10, 5$\times10^{-12}$), (20, 4$\times10^{-28}$) and (50, $10^{-90}$), respectively. Transition from $n_s < 1$ to $n_s > 1$ takes place at smaller values of $\lambda$ in comparison to \cite{Barrow}.}\label{fig:1}
\end{figure}

\begin{figure}[h]
\centering
\includegraphics[width=0.48\textwidth]{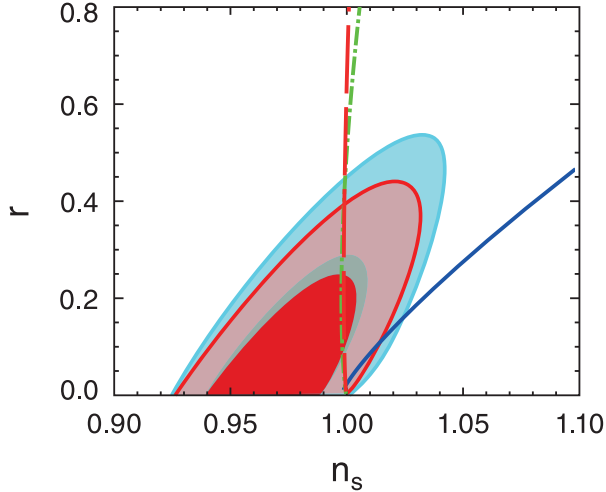}
\caption{The trajectories $r-n_s$ for different combinations of ($\lambda$, A). They have been compared with the five-years (blue regions) and seven years (red regions) data set of WMAP. In each case the contours show 68\% and 95\% confidence regions \cite{Larson}. The dash-dot green and long dash red curves which indicate a nearly Harrison-Zel'dovich model are related to the combinations (20, 2$\times10^{-28}$) and (50, $5\times10^{-92}$), respectively. The solid blue line, has been plotted for the combination (60, 3$\times10^{-109}$) and shows an upper bound for $\lambda$ in which the model is exiting the observational data. For larger values of $\lambda$ the model will be consistent with the data if we decrease enough other model parameter A.}\label{fig:2}
\end{figure}

Although we could not obtain a straight relation between $r$ and $n_s$, we can numerically illustrate some trajectories in the $r-n_s$ plane, if we fix our model parameters $\lambda$ and A. Since in logamediate inflationary model we only have a lower limit for $\lambda$, so we chose the values $\lambda = 2, 10, 20, 50$ to have a general comparison with the work \cite{Barrow}.
In figure (\ref{fig:1}), we have plotted four curves related to these values of $\lambda$ where in each case we have fixed the second model parameter A arbitrarily as they satisfy the condition (\ref{Alambda}). The solid yellow, dash black, dash-dot green and long dash red curves are related to the combinations (2, 10$^{-1}$), (10, 5$\times10^{-12}$), (20, 4$\times10^{-28}$) and (50, $10^{-90}$), respectively. It appears that the main difference between using a standard scalar field in a logamediate inflationary model \cite{Barrow} and a tachyonic field in it, is that in the latter, transition from $n_s<1$ to $n_s>1$ starts at smaller values of $\lambda$ in comparison to the former. We should mention that these curves have been plotted for as large as possible values of A satisfying (\ref{Alambda}) and if we choose some smaller values, then the curves move to the left. Thus, for the cases with $n_s > 1$ we can find some combinations of ($\lambda$, A) that in which the curves behave as a Harrison-Zel'dovich spectrum, i.e. $n_s = 1$. In figure (\ref{fig:2}), the dash-dot green and long dash red curves are related to the combinations (20, 2$\times10^{-28}$) and (50, 5$\times10^{-92}$), respectively.

Also, in figures (\ref{fig:1}) and (\ref{fig:2}) the trajectories have been compared with 68\% and 95\% confidence regions from observational data, i.e. five-years (blue-contours) and seven-years (red-contours) WMAP data set, which have been defined at $k_0=0.002$ Mpc$^{-1}$ \cite{Larson}. According to these observational data, an upper limit for $r$ has been found. This upper bound from five-years WMAP data set is $r<0.43$ whereas for seven-years data a stronger limit has obtained as $r<0.36$. In figure (\ref{fig:1}), the trajectories related to the combinations (2, 10$^{-1}$), (10, 5$\times10^{-12}$), (20, 4$\times10^{-28}$) and (50, $10^{-90}$), enter seven-years 95\% confidence region at $r\simeq0.25, 0.28, 0.39$ and 0.43, respectively. On the other hand, we can obtain the number of e-folds related to each one of these values of $r$. One can do this work numerically by first calculating when $\varepsilon=1$ which is the condition of beginning of inflation in our model and then replacing it in (\ref{nume-fold}). The resulting equation with (\ref{r}) gives the values of $N\simeq10175, 65, 16$ and 13, respectively. These values are proportional to the time spent by the tachyon field in the region of the $r-n_s$ plane allowed by the data and in each case our model is viable for larger values of the related $N$.

The solid blue line in figure (\ref{fig:2}) indicates the case in which we have considered the combination (60, 3$\times10^{-109}$) that it means for $\lambda>60$, the model tends to exit our observational contours unless we decrease the value of A much more.

\begin{figure}[h]
\centering
\includegraphics[width=0.48\textwidth]{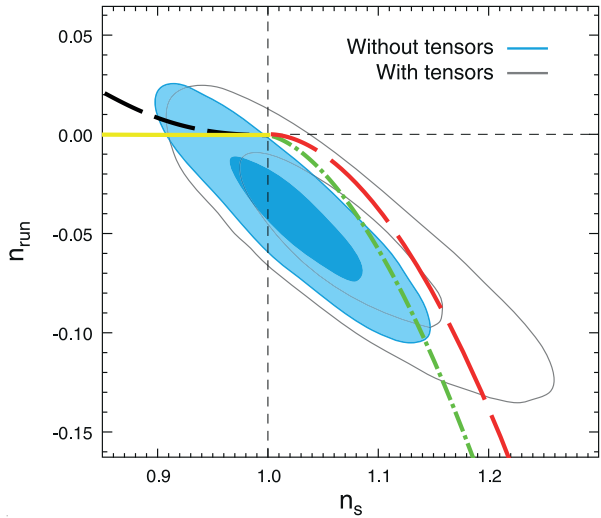}
\caption{The trajectories $n_{run}-n_s$ for different combinations of ($\lambda$, A). They have been compared with the five-years WMAP data set in two cases with and without considering tensor contributions. In each case the contours show 68\% and 95\% confidence regions. The solid yellow, dash black, dash-dot green and long dash red curves are related to the combinations (2, 10$^{-1}$), (10, 5$\times10^{-12}$), (20, 4$\times10^{-28}$) and (50, $10^{-90}$), respectively.}\label{fig:3}
\end{figure}

Figure (\ref{fig:3}) shows the dependence of the running of the scalar spectral index on the scalar spectral index parameter to lowest order for some combinations. Again the solid yellow, dash black, dash-dot green and long dash red curves are related to the combinations (2, 10$^{-1}$), (10, 5$\times10^{-12}$), (20, 4$\times10^{-28}$) and (50, $10^{-90}$), respectively. These curves have been compared with the contour plots from seven-years WMAP data set \cite{Larson} in which the negative values have been preferred. Seven-years data set implies that in models with only scalar fluctuations the marginalized value for the parameter $n_{run}$ is approximately -0.034 \cite{Larson},\cite{Komatsu}.

Also, it is obvious from this figure that for the combination (2, 10$^{-1}$) the model does not show any running in the scalar spectral index, at least in the lowest order.

\section{Is the model realistic?}

As noted in introduction, tachyonic potential has a special behavior. It starts from an unstable maximum at $\phi\rightarrow0$ and along its evolution $\frac{dV}{d\phi}<0$ until it approaches zero when $\phi\rightarrow\infty$. These are some motivations from string theory. To see how well-motivated is our model potential in (\ref{effv(t)}), first we derive $\dot H$ from (\ref{scale}) as
\begin{equation}\label{Hdot}
    \dot H = A\lambda(\ln t)^{\lambda-1}t^{-2}[\frac{\lambda-1}{\ln t}-1].
\end{equation}
Now, using (\ref{phidot}) we obtain
\begin{equation}\label{phidot2}
    \dot\phi^2=\frac{2}{3A\lambda}(\ln t)^{1-\lambda}(1+\frac{1-\lambda}{\ln t})
\end{equation}
At late times, one can neglect the second term in parenthesis above and consider $\dot\phi=\sqrt{\frac{2}{3A\lambda}}(\ln t)^{\frac{1-\lambda}{2}}$. But, integrating does not give us a good result. Assuming $\alpha=\sqrt{\frac{2}{3A\lambda}}$ and $\beta=\frac{1-\lambda}{2}$ one can obtain
\begin{equation}\label{phit}
    \phi-\phi_0=\alpha[t(\ln t)^\beta-\beta t(\ln t)^{\beta-1}+\beta(\beta-1)t(\ln t)^{\beta-2}-\beta(\beta-1)(\beta-2)t(\ln t)^{\beta-3}+...]
\end{equation}
and using this, we can not get $V(\phi)$, explicitly.

In another approach, assuming $\dot\phi^2 = \dot\Phi$ in (\ref{phidot2}), one can integrate and reach to the explicit function
\begin{equation}\label{Phit}
    \Phi(t)=\frac{2}{3A\lambda}t(\ln t)^{1-\lambda}
\end{equation}
and substituting this into (\ref{effv(t)}) give us $V(\Phi)=\frac{4}{3}\Phi^{-2}$. Although this form of potential is of interest but the thing we need is behavior of $V(\phi)$ and since we can not relate two functions $\phi(t)$ and $\Phi(t)$ so we can not establish $V(\phi)$ even approximately.

\begin{figure}[h]
\centering
\includegraphics[width=0.48\textwidth]{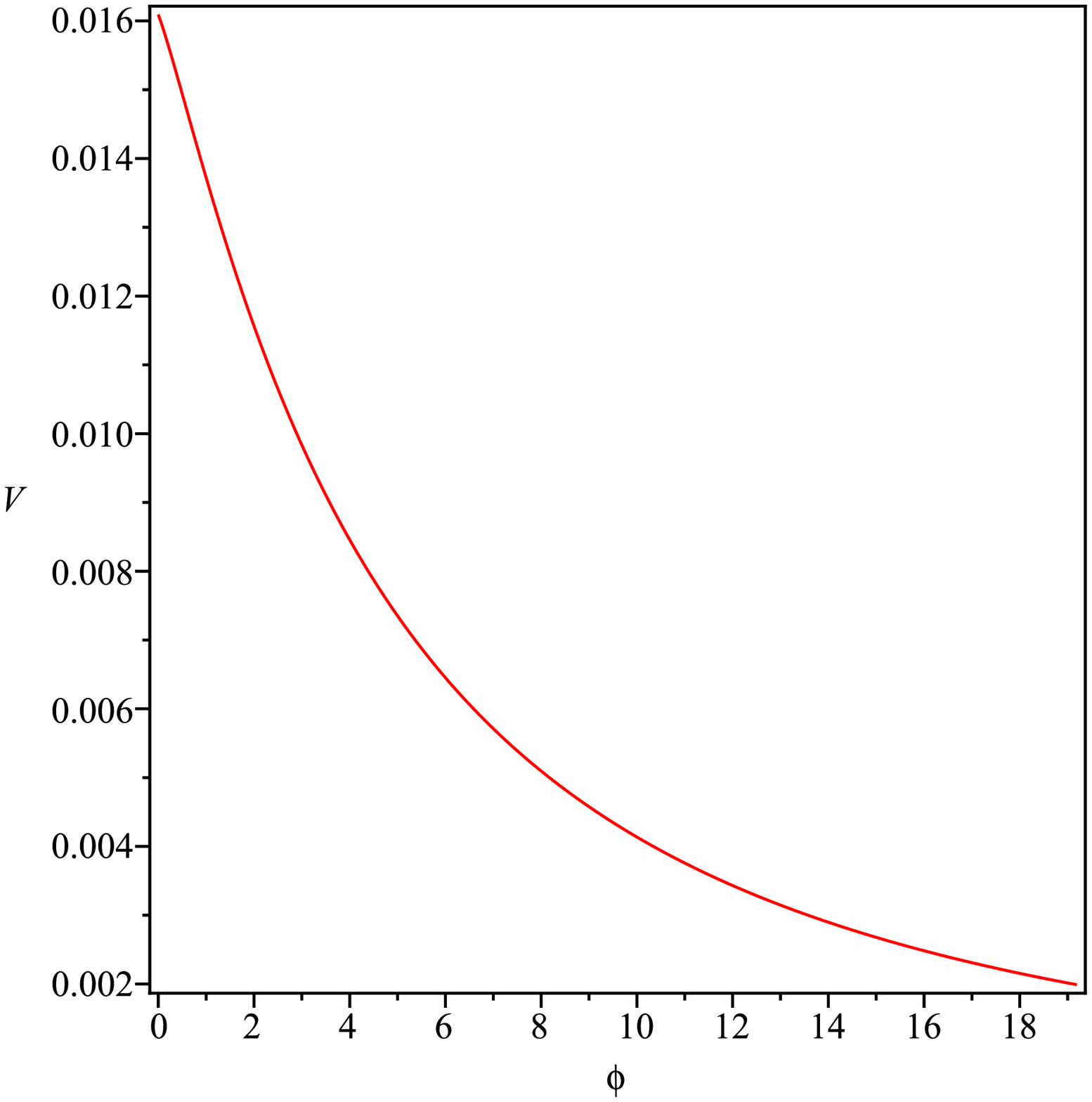}
\includegraphics[width=0.48\textwidth]{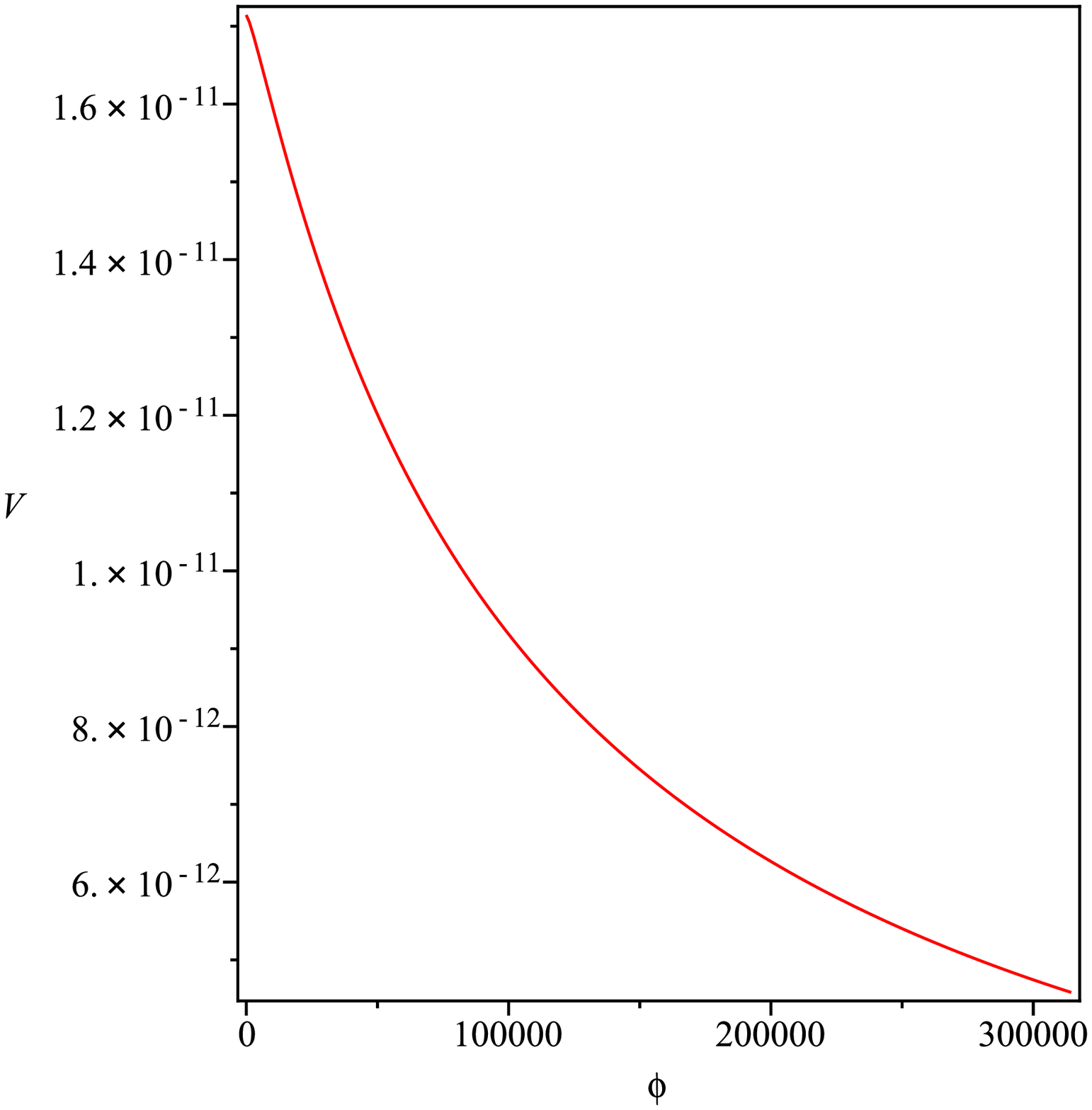}
\includegraphics[width=0.48\textwidth]{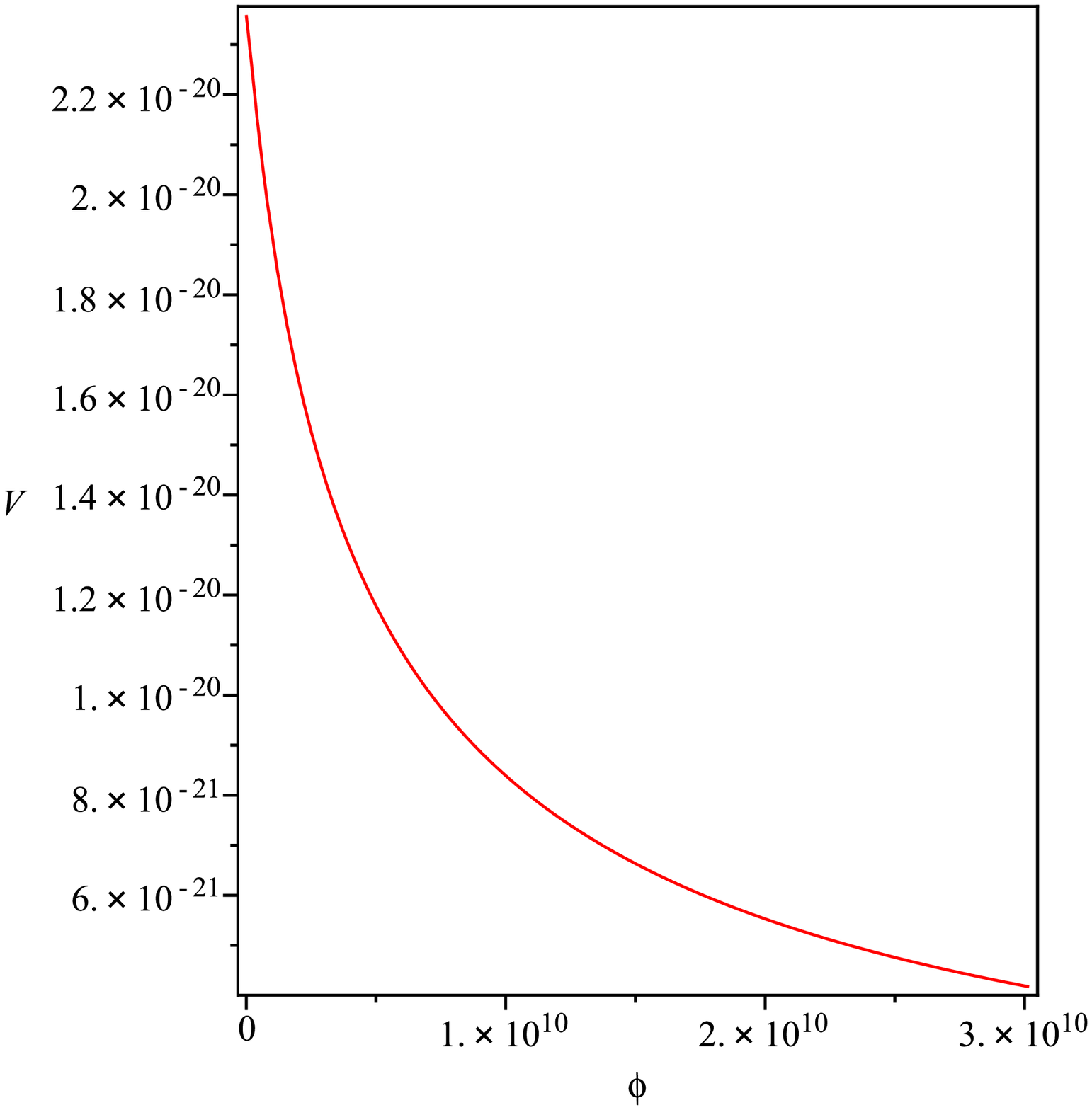}
\includegraphics[width=0.48\textwidth]{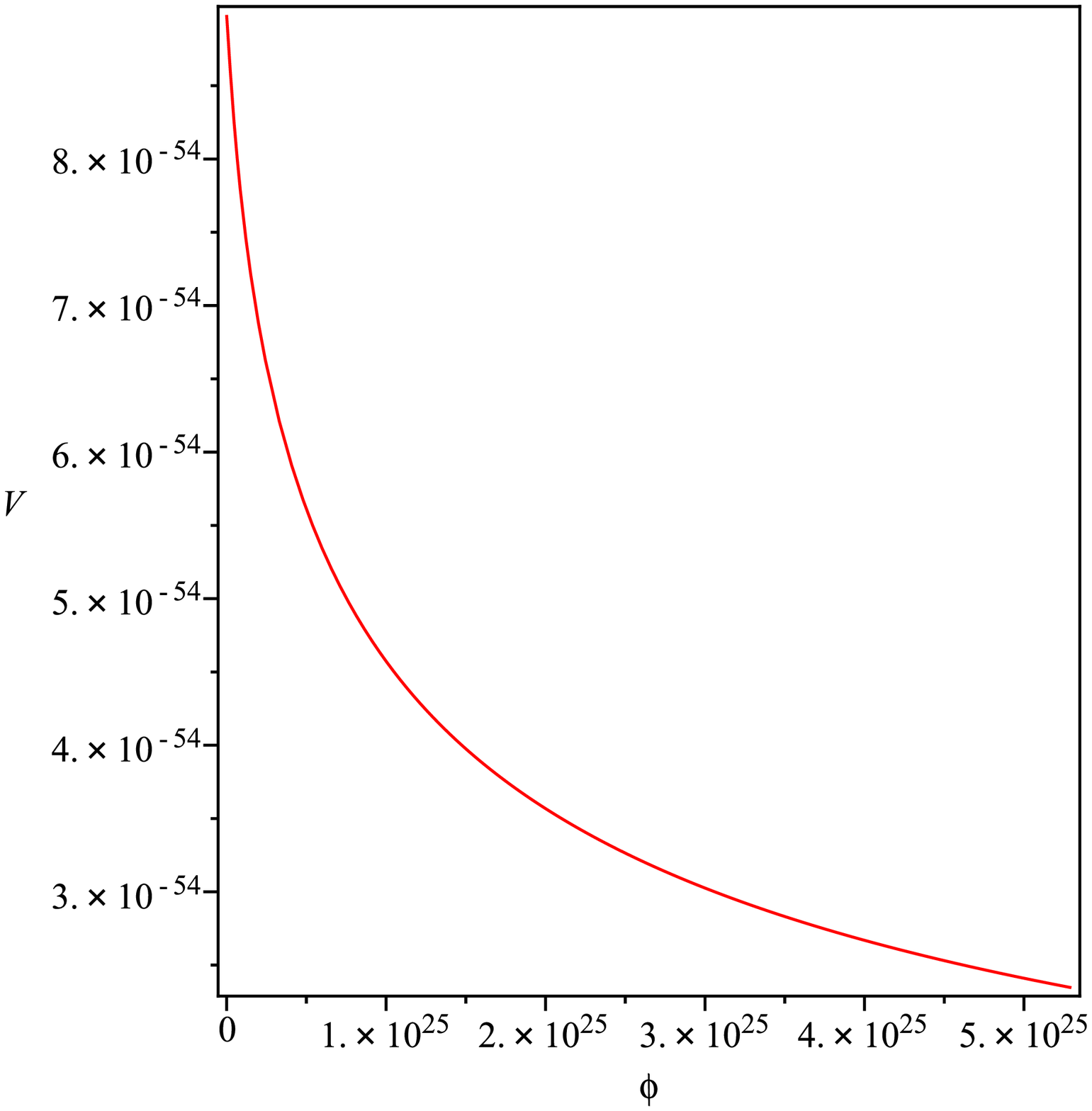}
\caption{The trajectories $V(\phi)$ for different combinations of ($\lambda$, A), up-left:(2, 10$^{-1}$), up-right:(10, 5$\times10^{-12}$), bottom-left:(20, 4$\times10^{-28}$) and right:(50, 10$^{-90}$). All necessary conditions for a tachyonic potential have been satisfied}\label{fig:4}
\end{figure}

So, we chose numerical approach to show how our model is realistic. We can do this using (\ref{Phit}), $V(\Phi)=\frac{4}{3}\Phi^{-2}$ and $\dot\Phi = \dot\phi^2$. Also, to do this we should fix some of our model parameters such as $A$ and $\lambda$. In figure (\ref{fig:4}) we have shown $V(\phi)$ for all combinations of $(\lambda, A)$ which we have used in figures (\ref{fig:1}) and (\ref{fig:3}), i.e. (2, 10$^{-1}$), (10, 5$\times10^{-12}$), (20, 4$\times10^{-28}$) and (50, 10$^{-90}$). It is obvious from these plots that the general conditions for the string theory tachyon field which noted above are satisfied. Indeed, we can say that our model potential is a well-motivated tachyon potential and the model under consideration is realistic. Also, we can see that increasing in $\lambda$ leads to more smooth behavior of $V(\phi)$.

\section{CONCLUSION}

In this work we studied the logamediate inflation in the presence of tachyon field. In the slow-roll approximation we derived the effective tachyon potential and the slow-roll parameters. Also, the number of e-folds which indicates how long inflation takes, was obtained.

Starting with a perturbed line element we investigated our model in the context of perturbation theory. We calculated some important parameters such as scalar spectral index $n_s$, its running $n_{run}$ and tensor to scalar ratio $r$ in our model. Then, we plotted some curves for different combinations of our model parameters ($\lambda$, A) and compared them with some observational data. From graph $r-n_s$ we concluded that our model is in a good agreement with observations for different combinations of the model parameters such as from (2, 10$^{-1}$) to (60, 3$\times10^{-109}$). Also, one can find some combinations that result Harrison-Zel'dovich spectrum, i.e. $n_s \simeq 1$, for example (20, 2$\times10^{-28}$) and (50, $5\times10^{-92}$).

In the last section we investigated whether our model and specially the resulting tachyonic potential is realistic or not. We could not do this analytically but numerical discussion was useful. In admissible combinations of $(\lambda, A)$ which we have used in this article, we could show that the general conditions of a tachyon potential are satisfied in our model.

\end{document}